\documentstyle[12pt]{article}
\newcommand{\be}{\begin{equation}}
\newcommand{\ee}{\end{equation}}
\newcommand{\ec}{\end{center}}
\newcommand{\bc}{\begin{center}}
\newcommand{\fr}{\frac}

\newcommand{\Lam}{\Lambda}

\begin{document}
\bc
{\Large  Cosmological Implications of The Bulk  Viscous Model With
Variable $G$ And $\Lam$}
\ec
\vspace{1cm}
\begin{center}
 ARBAB I. ARBAB  \\
\vspace{1cm}
{\it Department of Physics, Faculty of Science, University of
Khartoum,\\
P.O. 321, Khartoum 11115, Sudan.}\\
\end{center}
\vspace{3cm}
\bc
{\bf ABSTRACT}
\ec
We have investigated the cosmological implications of the bulk viscous
cosmological model with variable $G$ and $\Lam$. These results are found
to be compatible with the present observations. The classical cosmological
tests for this model encompass the Freese {\it et al.} ones.
The model has some spirits of the Standard Model.
 The inflationary solution which solves the Standard Model problems is
 obtained as a special solution.
The influence of viscosity is shown to affect the past and the future of
the Universe.

\vspace{1cm}

\newpage
\bc
{1. \bf Introduction}
\ec
We have recently discussed a bulk viscous model with variable $G$
and $\Lam$ in a scenario which conserves energy and momentum $[1]$.
We have found that the model has many interesting features. Various
models in the present literature are shown to be equivalent to bulk
viscous model. The bulk viscosity coefficient ($\eta$) is usually taken
to depend on the energy density ($\rho$) of the cosmic fluid, viz.
$\eta=\eta_0\rho^n$, where $\eta_0$ is a constant and $n$ is the viscosity `index'.
It is shown that only models with $n=1/2$ are structurally stable.
Beesham has shown that the Berman model is equivalent to a bulk viscous model.
We, however, have  shown that this equivalence is quite general.

The introduction of bulk viscosity is usually accompanied with particles
creation, since this term appears as an energy source in the energy conservation
equation. It is also shown that the introduction of a variable cosmological
constant solves the entropy problem. In this model neither of the two cases appear.
This can be understood as the fact that the conservation of energy in cosmology
is very restrictive. The conspiracy between the gravitational constant ($G$), cosmological
constant $(\Lambda$) and the bulk viscosity is required to accomplish this.
These three parameters are found to be related by a new  equation.

This model shares some spirits of the Standard Model on one hand and the
inflationary model on the other hand. But, in this case the inflation is
caused by viscosity effects.
The notable feature of this model is the fact that the Standard and the
inflationary Models are augmented in one model, and that the inflationary solution
does not require the Universe to have the critical energy density.\\
It is interesting to note that the inflationary solution can be
obtained with or without the cosmological constant. Also note that this
solution in this model requires $\eta=\eta_0\rho$ and
$H\propto\eta_0^{-1}$, where $H$ is the Hubble's constant.
So the smaller is the value of $\eta_0$ the bigger is the inflation rate.
This solution has been obtained by Murphy (1973) where he
attributed the viscosity effect to gravitons production in the graviton-graviton scattering.\\
The physical existence of this bulk viscosity is still needed to be studied.
On the other hand, if it has been proved that the gravitational constant is
varying with cosmic time, then this cosmology becomes the next candidate.

A variation of $G$ has many interesting consequences both in geology and astrophysics.
Canuto and Narlikar [9] have shown that the $G$ varying cosmology is consistent with
whatsoever cosmological observations presently available.

In this paper we would like to discuss the cosmological implications of this
model. Recently, Waga $[2]$ has discussed these implications in  decaying
vacuum energy flat models but with $G$ constant.
We have shown that the model of Freese {\it et al}. can be reproduced  from this model.
As for  Freese {\it et al}, the nucleosynthesis constraints restrict $\beta$ to be
$\beta < 0.1$.
Complying with the recent values of the age of the Universe our model
fits very well.
\newpage
\bc
{2. \bf The Model and Field equations}
\ec
In a Robertson Walker universe
\be
d\tau^{2}= dt^{2}- 
R^{2}(t)[\frac{dr^{2}}{1-kr^2}+r^{2}(d\theta^{2}+\sin\theta^{2}d\phi^{2})]
\ee
where $k$ is the curvature index.\\
Einstein's field equations with time dependent cosmological and gravitational
``constants''\\
\be
R_{\mu\nu}-\frac{1}{2}g_{\mu\nu}R=8\pi GT_{\mu\nu}+ \Lambda g_{\mu\nu}
\ee
and the perfect fluid energy-momentum tensor \\
\be
T_{\mu\nu}=(\rho+p)u_{\mu}u_{\nu}-pg_{\mu\nu}
\ee
, in the presence of bulk viscosity, give $[1]$
\be
3(p^*+\rho)\dot{R}=-(\frac{\dot{G}}{G}\rho+\dot{\rho}+\frac{\dot{\Lambda}}{8\pi G})R.
\ee
where $p^*=p-3\eta H$ and $\eta$ is the coefficient of viscosity.
The conservation of energy requires
\be
\dot{\rho}+3H(\rho+p)=0\ .
\ee
The pressure $(p)$ and the energy density $(\rho)$ are related by the equation of the state
\be
p=(\gamma-1)\rho\ \ , \gamma=\rm const.\ .
\ee
Therefore, eq.(4) yields
\be
9\eta H\dot{R}=(\frac{\dot{G}}{G}\rho+\frac{\dot{\Lambda}}{8\pi G})R
\ee
We take the  ansatz [6,8]
\be
\Lambda =3\beta H^{2},\ \ \ \beta \ \ \ \rm const.    .
\ee
Lima {\it et al.} argued in favor of a term of the type $\Lambda\sim H^2$.
They assumed, on dimensional grounds, that $\Lambda\sim\frac{1}{t^2_{Pl}}(\frac{t_{Pl}}{t_H})^n$
where $t_H$ is the Hubble's time, $t_{Pl}$ and $\ell_{Pl}$ are respectively the Planck's time and length.
They take $n=2$ in order to get rid of the $\hbar$ dependence, since the gravitational
effects today  are described by some classical theory of gravity. Therefore,
since $t_H\sim H^{-1}$ they concluded that $\Lambda\propto H^2$.
On the other hand, the constancy of the parameter $\beta=\rho_v/(\rho+\rho_v)$\  ($\rho_v$ is the vacuum energy density)
proposed by Freese {\it et al.} is equivalent, at the level of the Einstein's equations,
to considering $\Lambda=3\beta H^2$ too.
Equation (7) can be written as
\be
9\eta \frac{H}{R}=\frac{G'}{G}\rho+\frac{\Lambda'}{8\pi G},
\ee
where a prime denotes derivative w.r.t scale factor $R$ while a dot is the
derivative w.r.t to cosmic time $t$. In what follows we will consider a flat universe, $k=0$\\
The density parameter $(\Omega$) is given by
\be
\Omega=\frac{8\pi G\rho}{3H^2}=(1-\beta)  \ .
\ee
The solution of eq.(9), using eqs.(5), (6), (8) and (10) yields
\be
R(t)=[3D\gamma(1-n)]^{[1/3\gamma(1-n)]}t^{[1/3\gamma(1-n)]}\ \ \ \ , \ D=\rm const..
\ee
The Hubble parameter is $ H_0=\frac{1}{3(1-n)}\frac{1}{t_0}$
and the deceleration parameter is $ q_0= -\frac{R\ddot{R}}{\dot{R}^2}|_0=2-3n$
, $0\le n\le 1$ ( the subscript 0 denotes a present day quantity).
This gives $-1\le q_0\le2$.

\bc
{3. \bf Lookback Time}
\ec
The radiation travel time  (or lookback time) $t-t_0$ for a photon emitted
by a source at instant $t$ and received at $t_0$ is
\be
t_0-t=\int_R^{R_0}\fr{dR}{\dot R}\ ,
\ee                                        
where $R_0$ is the present scale factor of the Universe.
Using eq.(11) one can write
\be
R=A't^{1/3(1-n)} \ \ A'=\rm const.       \ ,
\ee
it follows that 
\be
\fr{R_0}{R}=1+z=(\fr{t_0}{t})^{1/3(1-n)}\ ,
\ee
or
\be
t=(1+z)^{-3(1-n)}t_0\ .
\ee
Using eqs.(11) and (15) we can write
\be
t_0-t=\fr{1}{3(1-n)H_0}[1-(1+z)^{-3(1-n)}]\ ,
\ee
or
\be
H_0(t_0-t)=\fr{1}{3(1-n)}[1-(1+z)^{-3(1-n)}] \ .
\ee
For small $z$ one obtains
\be
H_0(t_0-t)=z-(2-\fr{3}{2}n)z^2+...
\ee
and with $q=2-3n$ this transform into
\be
H_0(t_0-t)=z-(1+q/2)z^2+...  \ \ ,
\ee
If we take $z\rightarrow \infty $ in eq.(17) we readily obtain
$H_0t_0=\fr{1}{3(1-n)} ; \ n=1/2$ gives the well-known Einstein-de Sitter (ES)
result, 
\be
H_0(t_0-t)=\fr{2}{3}[1-(1+z)^{-3/2}]\ .
\ee
We recall that the FRW flat models, as preferred by inflation, have $q=1/2$
and $\Omega=2q=1$, whereas the dynamical estimates yield $0.1 \le \Omega\le 0.4$ $[5]$.
Such a problem can be resolved in our scenario.
\bc
{4. \bf Proper Distance d(z)}
\ec
A photon emitted by a source with coordinate $r=r_1$ at time $t=t_1$
 and received at time $t_0$ by an observer located  at $r=0$ will follow 
 a null geodesic with ($\theta, \phi $)=const.\ . The proper distance between
 the source and the observer is given by
 \be
 d=R_0\int_R^{R_0}\fr{dR}{\dot RR}\ ,
 \ee
 \be
 r_1=\int_{t_1}^{t_0}\fr{dt}{R}=\fr{H_0^{-1}R_0^{-1}}{2-3n}[1-(1+z)^{3n-2}]\ ,
 \ee
 hence
 \be
 d=r_1R_0=\fr{H_0^{-1}}{2-3n}[1-(1+z)^{3n-2}]\ .
 \ee
 For small $z$ this becomes
 \be
 H_0d=z-\fr{2}{3}(1-n)z^2+... = z-\fr{1}{2}(1+q)z^2+...
 \ee
 Equation (23) gives the Freese {\it et al.} result for the proper
 distance if we set $\Omega=1-\beta=2-2n$ viz.,
 \be
 H_0d=\fr{2}{3\Omega-2}[1-(1+z)^{1-(3/2)\Omega} ]\ .
 \ee
 As remarked before the ES result is a special case of eq.(23).
 Equation (23) shows that $d$ is maximum for $n\rightarrow 1 $
 (de Sitter) and minimum for $n\rightarrow 1/2 $ (ES).
Now if $n<2/3$, at $t_0$, the observer cannot see any source beyond the
proper distance of the particle horizon $[2]$
\be
d(z=\infty)=\fr{H_0^{-1}}{2-3n}\ ,
\ee
and for $2/3<n<1, d(z=\infty)$ is always infinite.
\newpage
\bc
{5. \bf Luminosity Distance}
\ec
This generalizes the inverse-square law of the brightness in the static
Euclidean space to an expanding curved space $[2]$
\be
d_L=(\fr{L}{4\pi\ell})^{1/2}=r_1R_0(1+z)
\ee
where $L$ is the total energy emitted by the source per unit time, $\ell$
is the apparent luminosity of the object and $r_1$  is the coordinate
distance. For flat universe 
$d=r_1R_0(1+z)$\ \ and  therefore one can write
\be
d_L=d(1+z)\ .
\ee
Using eq.(23) one gets
\be
H_0d_L=\fr{1+z}{2-3n}[1-(1+z)^{3n-2}]\ .
\ee
For $n=1/2$ the above equation gives back the ES result,
\be
d_L=\fr{2}{H_0}[(1+z)-(1+z)^{1/2}]\ .
\ee
For small $z$ eq.(29) gives
\be
H_0d_L=z+\fr{1}{2}(1-q)z^2=z+\fr{1}{2}(3n-1)z^2+...
\ee
\bc
{6. \bf Angular Diameter}
\ec
The angular diameter of a light source  of proper distance $D$ at
$r=r_1$ and $t=t_1$, observed at  $r=0$ and $t=t_0$ is given by
\be
\delta=\fr{D}{r_1R_1}=\fr{D(1+z)^2}{d_L}\ .
\ee
The angular diameter distance ($d_A$) is defined as the ratio  of the 
source diameter to its angular diameter
\be
d_A=\fr{D}{\delta}=r_1R_1=d_L(1+z)^{-2}\ .
\ee
Applying eq.(29)  we obtain
\be
d_A=\fr{H_0^{-1}}{2-3n}\fr{[1-(1+z)^{3n-2}]}{[1+z]}\ .
\ee
The maximum of $d_A$ for our model occurs at 
\be
1+z_m=[3(1-n)]^{1/(2-3n)}
\ee
while for the Freese {\it et al.} occurs at $[2]$
\be
1+z_m=(\fr{3}{2}\Omega)^{2/(3\Omega-2)}.
\ee
For $n=1/2$ we recover the ES result, viz. $z_m=5/4$.
\bc
{7. \bf Number Counts}
\ec
The number of astronomical sources in the volume element $dV$  at coordinate
$r=r_1$ is given by
\be
dN=nR^3r_1^2d\Omega dr_1
\ee
where $n$ is the number density of the sources and 
\be
dV=R^3r_1^2d\Omega dr_1, \ \ d\Omega=\sin\theta d\theta d\phi
\ee
Since the number of these object is conserved, we can write
\be
dN=n_0R_0^3r_1^2d\Omega dr_1
\ee
or
\be
dN=n_0(r_1R_0)^2d\Omega\fr{dR_0r_1}{dz}dz
\ee
Upon using eq.(29) this transforms into
\be
dN=\fr{n_0}{(2-3n)^2\ H_0^3}\fr{[1-(1+z)^{3n-2}]^2}{[1+z]^{3-3n}}d\Omega dz\ .
\ee
For $n=1/2$ we recover the ES result, namely
\be
dN=\fr{4n_0}{H_0^3}\fr{[1-(1+z)^{-1/2}]^2}{[1+z]^{3/2}}d\Omega dz\ .
\ee
This has a maximum value at $z= [\fr{7-9n}{3-3n}]^{1/(2-3n)}-1 $
whereas  in the case of ES this occurs at  $z=16/9 $ ( i.e. when $n=1/2$).

Equations (17), (23), (29), (34) and (41) are our predictions for the
corresponding cosmological tests. They are in agreement with the results
of the Freese {\it et al.} provided that the substitution  $\Omega=1-\beta=2-2n$ is made.
According to the Freese {\it et al.} only models with $\beta<0.1$ are consistent
with the Standard Model nucleosynthesis constraints. Hence one obtains a restriction
on $n$, viz. $n<0.55$ in that epoch.

\bc
{8. \bf Concluding Remarks}
\ec
In this work we have discussed the cosmological implications of the bulk
viscosity model with variable $G$ and $\Lam$. The results for the cosmological
tests are compatible  with the present observations.
The model of the Freese {\it et al.} is retrieved from this model for a particular
choice of $n$. The Einstein - de Sitter results are obtained for the case $n=1/2$.
While the parameter $\beta$ is a free parameter in the Freese {\it et al.} model,
it is a measure of viscosity in our present model. These tests are found to depend
on $\beta$ (or $n$) only. The proposed model resembles the Standard Model in
some respect and gives rise to inflation for the case $n=1$.
The model shows the importance of the influence of the bulk viscosity in the
evolution of the Universe.
Further studies are needed to account for the existence of this bulk viscosity.
\bc
{\bf References}
\ec
1- Arbab, A.I. {\it Gen. Rel. Gravit.29},61(1997)\\
2- Waga, I., {\it Astrophy. J. 414}, 436, 1993\\
3- Lima, J.A.S., $\&$ Maia, J.M.F., {\it Mod. Phy. Lett.A, 8}, 591 (1993)\\
4- Beesham, A. {\it Phy. Rev.D48,} 3539, 1993\\
5- Peebles, P.J.E, {\it Nature 321}, 27 (1986)\\
6- K.Freese, F.C.Adams, J.A.Frieman and E. Mottola {\it Nuc.Phys.B287},
797(1987)\\
7- G.L.Murphy, {\it Phys. Rev.D8}, 4231(1973)\\
8- Lima Carvalho Waga {\it Gen. Rel. Gravit.26}, 909(1994)\\
9- V.M. Canuto  and J.N. Narlikar {\it Astroph. J}.,6, 236(1980)\\
\end{document}